\documentclass[12pt]{article}
           
\usepackage{latexsym}
\usepackage{amssymb}
\usepackage{amsmath}
\usepackage{yhmath}

\def\im{\mathrm{i}}
\def\d{{\rm d}}
\renewcommand{\Re}{\operatorname{\mathrm{Re}}}
\renewcommand{\Im}{\operatorname{\mathrm{Im}}}

\newcommand{\G}[3]{G(#1;\, #2,#3)}
\def\Gz{G(z)}

\newcommand{\Gc}[3]{G^{*}(#1;\, #2,#3)}
\newcommand{\iG}[3]{(\Im G)(#1;\, #2,#3)}

\newcommand{\diag}[1]{\mathrm{diag}(#1)}
\newcommand{\abs}[1]{\ensuremath{|#1|}}

\def\Z{\mathbb Z}

\def\R{\mathbb R}
\def\C{\mathbb C}
\def\E{\mathbb E}

\newtheorem{thm}{Theorem}
\newtheorem{lem}[thm]{Lemma}

\title{A remark on an estimate by Minami}
\author{G.M. Graf, A. Vaghi\\
\normalsize\it Theoretische Physik,
ETH-H\"onggerberg, CH--8093 Z\"urich}

\begin{document}

\maketitle
\begin{abstract}
In the context of the Anderson model, Minami proved a Wegner type bound on the expectation of $2\times 2$ determinant of Green's functions. We generalize it so as to allow for a magnetic field, as well as to determinants of higher order. 
\end{abstract}

\bigskip
\section{Introduction}
Minami \cite{Mi} considered the Anderson model
\begin{equation*}
H=-\Delta + V
\end{equation*}
acting on $\ell^2(\Z^d)$, where $\Delta$ is the discrete Laplacian and $V=\{V_x\}_{x\in\Z^d}$ consists of independent, identically distributed real random variables, whose common density $\rho$ is bounded. He showed that in the localization regime the eigenvalues of the Hamiltonian restricted to a finite box $\Lambda\subset\Z^d$ are Poisson distributed if appropriately rescaled in the limit as $\Lambda$ grows large. The result and, up to small changes, its proof also apply when the kinetic energy $-\Delta$ is replaced by a more general operator $K=K^*$ with a rapid off-diagonal decay of its matrix elements $K(x,y)$ in the position basis ($x,y\in\Z^d$), as long as
\begin{equation}\label{eq:1}
K(x,y)=K(y,x)\, .
\end{equation}
Use of this property is made in the proof of Lemma 2 in \cite{Mi}, where $H$, and hence its resolvent $\Gz=(H-z)^{-1}$, is assumed symmetric: $\G{z}{x}{y}=\G{z}{y}{x}$, cf. eqs. (2.68, 2.75).\\

In physical terms, eq. (\ref{eq:1}) corresponds to the absence of an external
magnetic field, and it may thus be desirable to dispense with it. This is
achieved in this note. But first we recall Minami's Lemma 2. Let 
$\Im G=(G-G^*)/2\im$. Then \cite{Mi}
\begin{equation}\label{eq:2}
\E\Bigg[\det\left(\begin{array}{cc} 
\iG{z}{x}{x}& \iG{z}{x}{y}\\
\iG{z}{y}{x}& \iG{z}{y}{y}
\end{array}\right)\Bigg]\leq\pi^{2} \|\rho\|_{\infty}^{2}\, ,
\end{equation}
for $x\neq y$ and $\Im z>0$, and similarly if the Hamiltonian $H$ is truncated
to a subset $\Lambda\subset\Z^{d}$ with $x,y\in\Lambda$.

Because of $\Gc{z}{x}{y}=\overline{\G{z}{y}{x}}$, the above matrix element
\begin{equation}\label{eq:3}
\iG{z}{x}{y}=\frac{\G{z}{x}{y}-\overline{\G{z}{y}{x}}}{2\im}
\end{equation}
agrees with $\Im(\G{z}{x}{y})$ only if the symmetry (\ref{eq:1}) is assumed,
which we shall not do here. Then the agreement is limited to $x=y$. For the
sake of clarity we remark that it is the operator interpretation (\ref{eq:3})
of $\Im G$, and not the one in the sense of matrix elements, which makes
(\ref{eq:2}) true and useful in the general case.

The core of the argument is contained in the following
\begin{lem}\label{l1}
Let $A=(a_{ij})_{i,j=1,2}$ with $\Im A>0$. Then
\begin{multline}\label{eq:4}
\int \d v_1\d v_2\, \det\bigl(\Im[\diag{v_1,v_2}-A]^{-1}\bigr)\\
=\pi^{2}\frac{\det\Im A}{\sqrt{(\det\Im A)^{2}+\frac{1}{2}(\det\Im A)(\abs{a_{12}}^{2}+\abs{a_{21}}^{2})+\frac{1}{16}(\abs{a_{12}}^{2}-\abs{a_{21}}^{2})^{2}}}\, .
\end{multline}
\end{lem}

The right hand side is trivially bounded by $\pi^2$, since $\det\Im A>0$. In
\cite{Mi}, eqs. (2.72, 2.74), the equality was established in the special
case $a_{12}=a_{21}$. It was applied to
\begin{equation*}
-\bigl(A^{-1}\bigr)(u,v):=(\widehat{H}-z)^{-1}(u,v)\, ,\qquad(u,v=x,y)\, ,
\end{equation*}
where $\widehat{H}$ is $H$ with $V_x$ and $V_y$ set equal to zero. With the so
defined $2\times2$ matrix $A$ the two matrices under ``$\det\Im$'' in
(\ref{eq:2}) and on the left hand side of (\ref{eq:4}) agree, a fact known as
Krein's formula. That $A$ is actually well-defined and satisfies $\Im A>0$ is
seen from $\Im(z-\widehat{H})=\Im z>0$ and the following remarks \cite{Sie}, 
which apply to any complex $n\times n$ matrix $C$:
\begin{itemize}
\item[(i)] 
	\begin{equation}\label{eq:5}
	\Im C>0\;\Longleftrightarrow\;\Im(-C^{-1})>0\, .
	\end{equation}
	Indeed, $C$ is invertible, since otherwise $Cu=0$ for some $0\neq u\in\C^n$, implying 		$(u,(\Im C)u)=\Im(u,Cu)=0$, contrary to our assumption. Moreover, $\Im(-C^{-1})=C^{-1*}(\Im C)C^{-1}$. The converse implication is because $C\mapsto(-C)^{-1}$ is an involution.	
\item[(ii)] 
	\begin{equation}\label{eq:6}
	\Im C>0\;\Longrightarrow\;\Im\widehat{C}>0\, ,
	\end{equation}
	where $\widehat{C}$ is the restriction of $C$ to a subspace, as a sesquilinear form. In fact, $\Im\widehat{C}=\widehat{\Im C}$.
\end{itemize}
A more qualitative understanding of the bound $\pi^2$ for (\ref{eq:4}) may be obtained from its generalization to $n\times n$ matrices:
\begin{lem}\label{l2}
Let $A=(a_{ij})_{i,j=1,\dots,n}$ with $\Im A>0$. Then
\begin{equation*}
\int \d v_1\cdots\d v_n\, \det\bigl(\Im[\diag{v_1,\dots,v_n}-A]^{-1}\bigr)\leq
\pi^{n}\, .
\end{equation*}
\end{lem}

As a result, eq.~(\ref{eq:2}) also generalizes to the corresponding 
determinant of order $n$.

\section{Proofs}

{\bf Proof of Lemma \ref{l1}.} 
Following \cite{Mi} we will use that
\begin{equation}\label{eq:7}
\int\d x\, \frac{1}{\abs{ax+b}^2}=\frac{\pi}{\Im(\overline{b}a)}\,  ,\qquad(a,b\in\C,\, \Im(\overline{b}a)>0)
\end{equation}
and
\begin{equation}\label{eq:8}
\int\d x\, \frac{1}{ax^2+bx+c}=\frac{2\pi}{\sqrt{\Delta}}\,  ,\qquad(a>0,\,b,c\in\R,\, \Delta:=4ac-b^2>0)\,.
\end{equation}
We observe that
\begin{equation}\label{eq:8a}
\det\Im A=
(\Im a_{11})(\Im a_{22})-\frac{1}{4}\abs{a_{12}-\overline{a_{21}}}^2\, ,
\end{equation}
and hence the right hand side of (\ref{eq:4}), do not depend on $\Re a_{ii},\
(i=1,2)$. Similarly the left hand side, by a shift of integration variables. We
may thus assume $\Re a_{ii}=0$. The matrix on the left hand side of (\ref{eq:4}) is 
\begin{equation}\label{eq:9}
\Im\bigl[\diag{v_1,v_2}-A\bigr]^{-1}=\bigl(A^*-\diag{v_1,v_2}\bigr)^{-1}(\Im A)\bigl(A-\diag{v_1,v_2}\bigr)^{-1}\, ,
\end{equation}
and its determinant equals
\begin{equation}\label{eq:10}
\det\Im A\cdot\bigl|\det\bigl(A-\diag{v_1,v_2}\bigr)\bigr|^{-2}=
\det\Im A\cdot\bigl|(v_1-a_{11})(v_2-a_{22})-a_{12}a_{21}\bigr|^{-2}\,.
\end{equation}
The $v_2$-integration of the second factor (\ref{eq:10}) is of the type 
(\ref{eq:7}) with $a=v_1-a_{11}$ and $b=(a_{11}-v_1)a_{22}-a_{12}a_{21}$. Then
\begin{align}\label{eq:11}\nonumber
\Im(\overline{b}a)&=(\Im a_{22})\abs{v_1-a_{11}}^2 +\Im(a_{12}a_{21})\bigl(v_1-\Re a_{11}\bigr)+\Re(a_{12}a_{21})(\Im a_{11})\\
\nonumber
\\
&=(\Im a_{22})v_1^2+\Im(a_{12}a_{21})v_1+(\Im a_{22})(\Im a_{11})^2 +\Re(a_{12}a_{21})(\Im a_{11})\, .
\end{align}
By (\ref{eq:8}), the $v_1$-integral is obtained by computing the discriminant $\Delta$ of this quadratic function:
\begin{gather}\label{eq:12}
\int \d v_1\d v_2\, \bigl|\det\bigl(A-\diag{v_1,v_2}\bigr)\bigr|^{-2}=\frac{2\pi^2}{\sqrt{\Delta}}\, ,\\
\begin{aligned}
\\
\nonumber
\Delta &=4(\Im a_{11}\Im a_{22})^2+4(\Im a_{11}\Im a_{22})\Re(a_{12}a_{21})-\bigl(\Im(a_{12}a_{21})\bigr)^2\\
&=\bigl(2 \Im a_{11}\Im a_{22}+\Re(a_{12}a_{21})\bigr)^2-\abs{a_{12}a_{21}}^2\, .
\end{aligned}
\end{gather}
\noindent
In doing so we tacitly assumed that $\Delta$, and hence (\ref{eq:11}), are positive. This is indeed so, because $\Delta\leq0$ would imply that $A-\diag{v_1,v_2}$ is singular for some $v_1,v_2\in\R$, which contradicts $\Im A>0$, cf. (\ref{eq:5}). It also follows because $\Delta/4$ equals the expression under the root in (\ref{eq:4}), a claim we need to show anyhow: from (\ref{eq:8a}) and 
\begin{equation*}
\abs{a_{12}-\overline{a_{21}}}^2=\abs{a_{12}}^2+\abs{a_{21}}^2-2 \Re(a_{12}a_{21})
\end{equation*}
we obtain
\begin{equation*}
2 \Im a_{11}\Im a_{22}+\Re(a_{12}a_{21})=2\det\Im A+\frac{1}{2}\bigl(\abs{a_{12}}^2+\abs{a_{21}}^2\bigr)
\end{equation*}
and hence
\begin{equation*}
\Delta=4(\det\Im A)^2+2\bigl(\abs{a_{12}}^2+\abs{a_{21}}^2\bigr)(\det\Im A)+
\frac{1}{4}\bigl(\abs{a_{12}}^2+\abs{a_{21}}^2\bigr)^2-\abs{a_{12}a_{21}}^2\, .
\end{equation*}
Since the last two terms equal $(\abs{a_{12}}^2-\abs{a_{21}}^2)^{2}/4$, we
establish the claim and, by eqs.~(\ref{eq:10}, \ref{eq:12}), the lemma.
\hfill$\square$

\medskip\noindent
{\bf Proof of Lemma \ref{l2}.} 
By induction in $n$. It may start with $n=0$, in which case the determinant is
$1$ by natural convention, or with $n=1$, where the claim, i.e.,  
\begin{equation*}
\int\d v\, \Im\Bigl(\frac{1}{v-a}\Bigr)\leq\pi\, ,\qquad(\Im a >0)\, ,
\end{equation*}
is easily seen to hold as an equality. We maintain the induction step
\begin{equation*}
\int\d v_{n}\, \det\bigl(\Im[\diag{v_1,\dots,v_n}-A]^{-1}\bigr)\leq
\pi\det\bigl(\Im[\diag{v_1,\dots,v_{n-1}}-B]^{-1}\bigr)
\end{equation*}
for some $(n-1)\times(n-1)$ matrix $B$ with $\Im B>0$. This is actually a
special case of 
\begin{equation}\label{eq:13}
\int\d v\, \det\bigl(\Im[\diag{0,\dots ,0,v}-A]^{-1}\bigr)\leq
\pi\det\Im(-B)^{-1}\, ,
\end{equation}
where $B$ is the Schur complement of $a_{nn}$, given as 
\begin{equation}\label{eq:13a}
B=\widehat{A}-a_{nn}^{-1}(a_V\otimes a_H)
\end{equation}
in terms of the $(n-1,1)$-block decomposition of an $n\times n$ matrix:
\begin{equation*}
C=\left(\begin{array}{cc} \widehat{C}& c_V\\ c_H&c_{nn}\end{array}\right)\, .
\end{equation*}
 
By a computation similar to (\ref{eq:9}) the integrand in (\ref{eq:13}) is
\begin{multline*}
\frac{\det\Im A}{\bigl|\det\bigl(A-\diag{0,\dots ,0,v}\bigr)\bigr|^2}=\frac{\det\Im A}{\bigl|\det A-v\det\widehat{A}\bigr|^2}\\
=\frac{\det\Im A}{|\det A|^2\bigl|1-v(A^{-1})_{nn}\bigr|^2}
=\frac{\det\Im(-A^{-1})}{\bigl|1-v(A^{-1})_{nn}\bigr|^2}\, .
\end{multline*}
In the first line we used that $v\in\R$ and that the determinant is linear in the last row; in the second that 
\begin{equation}\label{eq:14}
(C^{-1})_{nn}\cdot\det C=\det\widehat{C}\, .
\end{equation}
By (\ref{eq:7}) the integral is $\pi$ times 
\begin{equation*}
\frac{\det\Im(-A^{-1})}{\Im(-A^{-1})_{nn}}=
\frac{\det\Im(-A^{-1})}{\bigl(\Im(-A^{-1})\bigr)_{nn}}\leq
\det\bigl[\widehat{\Im(-A^{-1})}\bigr]=\det\Im\bigl[-\widehat{A^{-1}}\bigr]\, ,
\end{equation*}
where the estimate is by applying 
\begin{equation*}
\det C\leq c_{nn}\cdot\det\widehat{C}\, ,\qquad(C>0)
\end{equation*}
to $C=\Im(-A^{-1})$, cf. (\ref{eq:5}). This inequality is by Cauchy's for the sesquilinear form $C$: letting $\delta_n=(0,\dots,0,1)$,
\begin{equation*}
1=(C^{-1}\delta_n,C\delta_n)^2\leq
(C^{-1}\delta_n,CC^{-1}\delta_n)\cdot(\delta_n,C\delta_n)=
(C^{-1})_{nn}\cdot c_{nn}\, , 
\end{equation*}
cf. (\ref{eq:14}). Finally, $\widehat{A^{-1}}$ may be computed by means of the
Schur (or Feshbach) formula \cite{HoJo}: $\widehat{A^{-1}}=B^{-1}$ with $B$ 
as in (\ref{eq:13a}).
Note that the left hand side is invertible because of
$\Im(-\widehat{A^{-1}})>0$, cf. (\ref{eq:5}, \ref{eq:6}), and that it is 
the inverse of a matrix with positive imaginary part. 
\hfill$\square$

%\medskip\noindent
%{\bf Acknowledgements.} This contribution  

\end{document}